\documentclass[runningheads]{smtda}

\usepackage{graphicx} 
\usepackage{amsmath}
\usepackage{amssymb}
\usepackage{float, epstopdf}

\usepackage{times}
\usepackage{color}
\usepackage{soul}                    
\definecolor{red}{rgb}{1,0,0}
\definecolor{green}{rgb}{0,1,0}
\definecolor{blue}{rgb}{0,0,1}

\begin{document}
%
\title*{Modeling and analysis of cyclic inhomogeneous Markov processes: 
        a wind turbine case study}
%
\toctitle{Modeling and analysis of cyclic inhomogeneous Markov processes}
%
\titlerunning{Modeling and analysis of cyclic inhomogeneous Markov processes}
%
\author{
  Teresa Scholz\inst{1,2}
  \and
  Vitor V.~Lopes\inst{3,4}
  \and 
  Pedro G.~Lind\inst{5}
  \and
  Frank Raischel\inst{6,7}
}

%
\index{Scholz, T.}
\index{Lopes, V.V}
\index{Lind, P.G.}
\index{Raischel, F.}

%
\authorrunning{Scholz et al}
%
\institute{
  Center for Theoretical and Computational Physics, University of Lisbon, 
  Portugal
  \and
  Energy Analysis and Networks Unit, National Laboratory of Energy and 
  Geology, Lisbon, Portugal\\
  (e-mail: {\tt teresa.scholz@lneg.pt})
  \and
  DEIO-CIO, Science faculty, University of Lisbon, Portugal
  \and 
  Universidad de las Fuerzas Armadas-ESPE, Latacunga, Ecuador
  \and
  ForWind-Center for Wind Energy Research, Institute of Physics, Carl-von-Ossietzky University of Oldenburg, Oldenburg, Germany\\
  \and
  Department of Theoretical Physics, University of Debrecen, Debrecen, Hungary
  \and
  Center for Geophysics, IDL, University of Lisbon, Portugal  
}

\maketitle             

\begin{abstract}
   A method is proposed to reconstruct a cyclic time-inhomogeneous Markov process from measured data. First, a time-inhomogeneous Markov model is fit to the data, taken here from measurements on a wind turbine. From the time-dependent transition matrices, the time-dependent Kramers-Moyal coefficients of the corresponding stochastic process are computed. Further applications of this method are discussed. 
\end{abstract}

\section{Introduction}

    Many complex systems can be described, within a certain level of modelization, as stochastic processes. A general stochastic process can be characterized in the linear noise approximation through a Fokker-Planck equation, in  continuous variables. For dealing with discrete variables in discrete time steps, often Markov Chains are the models of choice. Although in many cases both approaches converge in the limit of small time steps and increments of the stochastic variables, this correspondence is in general non-trivial\cite{vankampen}. In the Fokker-Planck picture, the  so-called Kramers-Moyal (KM) coefficients provide a complete description of a given stochastic process\cite{risken}.

    In the past decades, numerical procedures have been established to estimate the KM coefficients from measured stochastic data, which are applicable for any stationary, i.e.~time-homogeneous, Markov process. These methods require large sequences of data, but they are robust\cite{friedrich11}, have well-known errors and limitations\cite{kleinhans12}, require little intervention and are typically parameter-free\cite{friedrich11,lind2010}.

    However, for non-stationary Markov processes, much fewer methods and results are available to our knowledge. In this case,  estimations of the time-dependent KM coefficients can be obtained by two approaches: either the data  from the  inhomogeneous process is split into shorter, homogeneous sequences, on which then an estimate  of the KM coefficients can be performed through the aforementioned  methods\cite{Mourik06}. Or, if the inhomogeneous process is also cyclic, a parametrized time-dependent {\it ansatz} for the KM coefficients can be fit to the data\cite{Micheletti2008}. Compared to the stationary processes, both approaches for the inhomogeneous case  require a much higher level of pre-analysis, guesswork and iterative improvement. 

    In this paper, we present a method that allows to estimate the transition matrices of a time-inhomogeneous Markov model from data. As reported in a previous publication\cite{Scholz2014}, this method provides results with a considerable level of accuracy. Under well-known limitations, the discrete Markov model corresponds  to a continuous stochastic process in the form of a Fokker-Planck equation, which is completely characterized, in this case, through its time-dependent KM coefficients.  From the transition matrices, we can immediately calculate these KM coefficients, and therefore characterize the dynamical features underlying the time-dependent stochastic process.            

    We apply this methodology to data from a turbine in a wind park, where measurements of the wind velocity and direction, and the electric power output of the turbine are taken in $10$ minute intervals. The  results presented from this analysis show the general applicability of our method and are in agreement with previous findings.

    This paper is organized as follows. We start in Sec.~\ref{sec:methods} by introducing both the cyclic time-dependent Markov model and the procedure for extracting stochastic evolution equations directly from data series. In Sec.~\ref{sec:data} we describe the data and in Sec.~\ref{sec:D1D2} we present the time-dependent functions that define the stochastic evolution of the state of the wind turbine. Section \ref{sec:conclusions} concludes the paper.

\section{Methodology}
\label{sec:methods}

    This section describes the methodology used for the data analysis. In Sec.~\ref{section:MarkovProcess} the cyclic inhomogeneous Markov model to represent the daily patterns in the data is described and in Sec.~\ref{section:Langevin} we explain how stochastic evolution equations are derived directly from the Markov process transition matrices.
    
    \subsection{Modelling cyclic time-dependent Markov processes}
    \label{section:MarkovProcess}
    
        The goal of this time-inhomogeneous Markov process is to get a model that accurately reproduces the long-term behavior while considering the daily patterns observed in the data. Thus, the proposed objective function combines two maximum likelihood estimators: the first term maximizes the likelihood of the cycle-average probability; and, the second term maximizes the likelihood of the time-dependent probability. The final optimization problem is transformed into a convex one using the negative logarithm of the objective function. This section gives a brief overview over the final optimization problem. A detailed description of the objective function, the parametrization of the time-variant probability functions, and the constraints that must be added to the optimization problem to ensure its Markov properties is provided in \cite{Scholz2014}.


        A discrete finite Markov process $\{X_t \in S, t \geq 0\}$ is a stochastic process on a discrete finite state space $S = \{s_1, ..., s_n\}$,~$n \in \mathbb{N}$, whose future evolution depends only on its current state \cite{Kemeny1976}. 

        It can be fully described by the conditional probability $Pr\{X_{t+1} = s_j\mid X_{t} = s_i\}$ of the Markov process moving to state $s_j$ at time step $t+1$ given that it is in state $s_i$ at time $t$. It is called the $t$-th step transition probability, denoted as $p_{i,j}(t)$. 
        
        Being time-dependent, the Markov process has associated transition probability matrices $P_t$ that change with time. Considering $n$ possible states, the matrices $P_t$ have dimension $n\times n$ with entries $[P_t]_{i,j} = p_{i,j}(t)$ for all $i,j = 1,\ldots, n$, satisfying $p_{i,j}(t) \geq 0$ and $\sum_{j}{p_{i,j}(t)} = 1$.

        Markov process is called cyclic with period $T \in \mathbb{R}$, if $T$ is the smallest number, such that $p_{i,j}(mT + r) = p_{i,j}(r)$ for all $m \in \mathbb{N}$ and $0 \leq r < T$. See Ref.~\cite{Platis1998}. Since this paper deals with discrete data, $T$ and $r$ can considered to be multiples of the time step $\Delta t$ between successive data points and therefore integers. One can describe the cyclic Markov process by $T$ transition matrices $P_r$,~$r = 0,...,T-1$. The remainder of time step $t$ modulo $T$ will be denoted as $r_t$ and consequently $r_t=r_{t+mT}$. We fix $T=1$ day and use $\Delta t =1$.


        In this paper, the transition probabilities $p_{i,j}(z)$ are modeled by Bernstein polynomials, namely
            \begin{equation}
                p_{i,j}(z) = \sum_{\mu = 0}^k \beta_{\mu}^{i,j}b_{\mu,k}(z),
            \end{equation}
        where $z=r/T$ indicates the time of the day ($T=1$ day), $b_{\mu,k}(z)$ is the $\mu$-th Bernstein basis polynomial of order $k$, and $\beta_{\mu}^{i,j} \in \mathbb{R}$. The choice of these polynomials has several advantages properly described in \cite{Scholz2014}.

        To maximize the likelihood of the time-dependent transition probabilities given the data, the objective function must consider the time of the day $z$ when the transition happens. The corresponding term of the objective function is thus given by $\sum_{(i,j)_z\in \mathcal{S}_z} \log(p_{i,j}(z))$, where $\mathcal{S}_z$ is the set of observed transitions together with the time $z$ when they happens. This estimator allows to compute the intra-cycle transition probability functions, and thus to represent the daily patterns present in the data.

        A second term is added to this function, namely $\sum_{(i,j)\in \mathcal{S}} \log(p^{avg}_{i,j})$, where $\mathcal{S}$ is the set of transitions observed in the data and $p_{i,j}^{avg}$ is the cycle-average (daily) probability of transition from state $s_i$ to $s_j$. It is given by        
        $p_{i,j}^{avg} = \frac{1}{k + 1} \sum_{\mu = 0}^k \beta^{i,j}_k$.        
        This second term is the maximum likelihood estimator for the daily average probability and its addition to the objective function increases the consistency of the long-term behavior of the Markov process with the data. 

        Using the resulting overall objective function the optimization problem to be solved for the transition probability coefficients $\beta_{\mu}^{i,j}$ is translated into the minimization of 

        \begin{equation}
        {\cal L} = - \sum_{(i,j)\in \mathcal{S}} \log(\frac{1}{k + 1} \sum_{\mu = 0}^k \beta^{i,j}_{\mu}) - \sum_{(i,j)_z\in \mathcal{S}_z} \log(\sum_{\mu = 0}^k \beta_{\mu}^{i,j}b_{\mu,k}(z))
        \end{equation}
        subject to
        \begin{subequations}
        \begin{eqnarray}
        \sum_{j} \beta_{\mu}^{i,j} &=& 1  \label{rowstochasticity} \\
        \beta_0^{i,j} &=& \beta_k^{i,j}  \label{c0} \\
        \beta_{0}^{i,j} &=& \tfrac{1}{2}(\beta_{1}^{i,j} + \beta_{k-1}^{i,j}) \label{c1} \\
        \beta^{i,j}(w) &\leq& 1 \label{bounds1}\\
        0 &\leq& \beta^{i,j}(w) \label{bounds2} 
        \end{eqnarray}
        \end{subequations}        
        with $i,j=1,\dots,n$ and $\mu=0,\dots,k$, $k$ being the order of the Bernstein polynomials and $w$ the number of subdivisions.
        Constraint (\ref{rowstochasticity}) assures the row-stochasticity of the transition matrices, while constraints (\ref{c0}) and (\ref{c1}) impose $\mathcal{C}^0$- and $\mathcal{C}^1$-continuity at $z = 0$. Constraints~(\ref{bounds1}) and~(\ref{bounds2}) bound the transition probabilities between $0$ and $1$. 
        These constraints are derived using a property of the Bernstein polynomials to always lie in the convex hull defined by their control points $ (\frac{k}{\mu}, \beta_{\mu})$,~$\mu = 0,...,k$. This convex hull bound can be tightened by subdivision using the de Casteljau algorithm. In the resulting constraints (\ref{bounds1}) and~(\ref{bounds2}), $w$ is the number of subdivisions.
    
    \subsection{Extracting the stochastic evolution equation}
    \label{section:Langevin}



        In this section we characterize general stochastic processes through a Fokker-Planck equation. We consider
        a $N$-dimensional stochastic process $\mathbf{X}=(X_1(t),\dots,X_N(t))$ whose probability density function (PDF) $f(\mathbf{X},t)$ evolves according to the Fokker-Planck equation (FPE) \cite{risken} 
        \begin{eqnarray}
        \frac{\partial f(\mathbf{X},t)}{\partial t} &=&
            -\sum_{i=1}^N\frac{\partial}{\partial x_i}
                \left [ 
                D_i^{(1)}(\mathbf{X})f(\mathbf{X,t})
                \right ] \cr
            & &+\sum_{i=1}^N\sum_{j=1}^N
                \frac{\partial ^2}{\partial x_i\partial x_j}
                \left [
                D_{ij}^{(2)}(\mathbf{X})f(\mathbf{X},t)
                \right ] \quad .
        \label{FPE}
        \end{eqnarray}
        The functions $D_i^{(1)}$ and $D_{ij}^{(2)}$ are the first and second Kramers-Moyal coefficients respectively, more commonly called the drift and diffusion coefficients. 

        These coefficients provide a complete description of a given stochastic process and are defined as
        \begin{equation}
        \mathbf{D}^{(k)}(\mathbf{X})=\lim_{\Delta t\rightarrow0}\frac{1}{\Delta t}
                                        \frac{\mathbf{M}^{(k)}(\mathbf{X},\Delta t)}{k!}
        \label{DefCoefKM}\quad,
        \end{equation}
        where $\mathbf{M}^{(k)}$ 
        are the first ($k=1$) and second ($k=2$) conditional moments. $\mathbf{D}^{(1)}$ is the drift vector and $\mathbf{D}^{(2)}$ 
        the diffusion matrix.

        If the  underlying process is stationary and therefore both drift and diffusion coefficients do not explicitly depend on time $t$, the conditional moments can be directly derived from the measured data as \cite{friedrich11,lind2010}:
        \begin{equation}
        \begin{array}{lcl}
        M_i^{(1)}(\mathbf{X},\Delta t) = 
            \left\langle Y_i(t+\Delta t)-Y_i(t)  | {\mathbf{Y}(t)=\mathbf{X}} \right\rangle  & & \\
        M_{ij}^{(2)}(\mathbf{X},\Delta t) = & &  \\
            \left\langle (Y_i(t+\Delta t)-Y_i(t))(Y_j(t+\Delta t)-Y_j(t))
            | {\mathbf{Y}(t)=\mathbf{X}}\right\rangle,  & & 
        \end{array}
        \label{2D}
        \end{equation}
        where $\mathbf{Y}(t)=(Y_1(t),\dots,Y_N(t))$ exhibits the $N$-dimensional vector of measured variables at time $t$ and $\langle \cdot
        | {\mathbf{Y}(t)=\mathbf{X}} \rangle$ symbolizes a conditional averaging over the entire measurement period, where only measurements with ${\mathbf{Y}(t)=\mathbf{X}}$ are taken into account.
        In practice binning or kernel based approaches with a certain threshold are applied in order to evaluate the condition $\mathbf{Y}(t)=\mathbf{X}$. See e.g.~Ref.~\cite{friedrich11} for details.

        If the process is non-stationary and time-inhomogeneous, we must consider an explicit time-dependence of the KM coefficients, 
        which translates into time-dependent conditional moments 
        that can be calculated using a short-time propagator\cite{friedrich11}.
        In our case, this short-time propagator corresponds to the transition probabilities $p_{i,j}(t)$, yielding for the conditional moments 
        
        \begin{eqnarray}
        M^{(l)} (P_k,v_k,\theta_k, t+\Delta t) &=&  \left( \begin{array}{c}  %
        \sum_{j}p_{k,j} (t) \left(v_{j}-v_k\right)^l \\ \sum_{j}p_{k,j}(t) \left(P_{j}-P_k\right)^l \\ \sum_{j}p_{k,j}(t)\left(\theta_{j}-\theta_k\right)^l  
        \end{array} \right) \, .
        \end{eqnarray}
        


\section{Data: wind and power at one wind turbine}
\label{sec:data}

    The data for this study was obtained from a wind power turbine in a wind park located in a mountainous region in Portugal. The time series consists of a three-year period (2009-2011) of historical data gotten from the turbine data logger. The sampling time of 10 minutes leads to 144 samples each day. The data-set comprises three variables, wind power, speed and direction (nacelle orientation). The wind speed information was collected from the anemometer placed in the wind turbine hub. Due to confidentiality, wind power and speed data values are reported as a fraction of the rated power and the cut-out speed, respectively.

\begin{figure}[H]
        \centering
        \includegraphics[width = 0.5\textwidth]{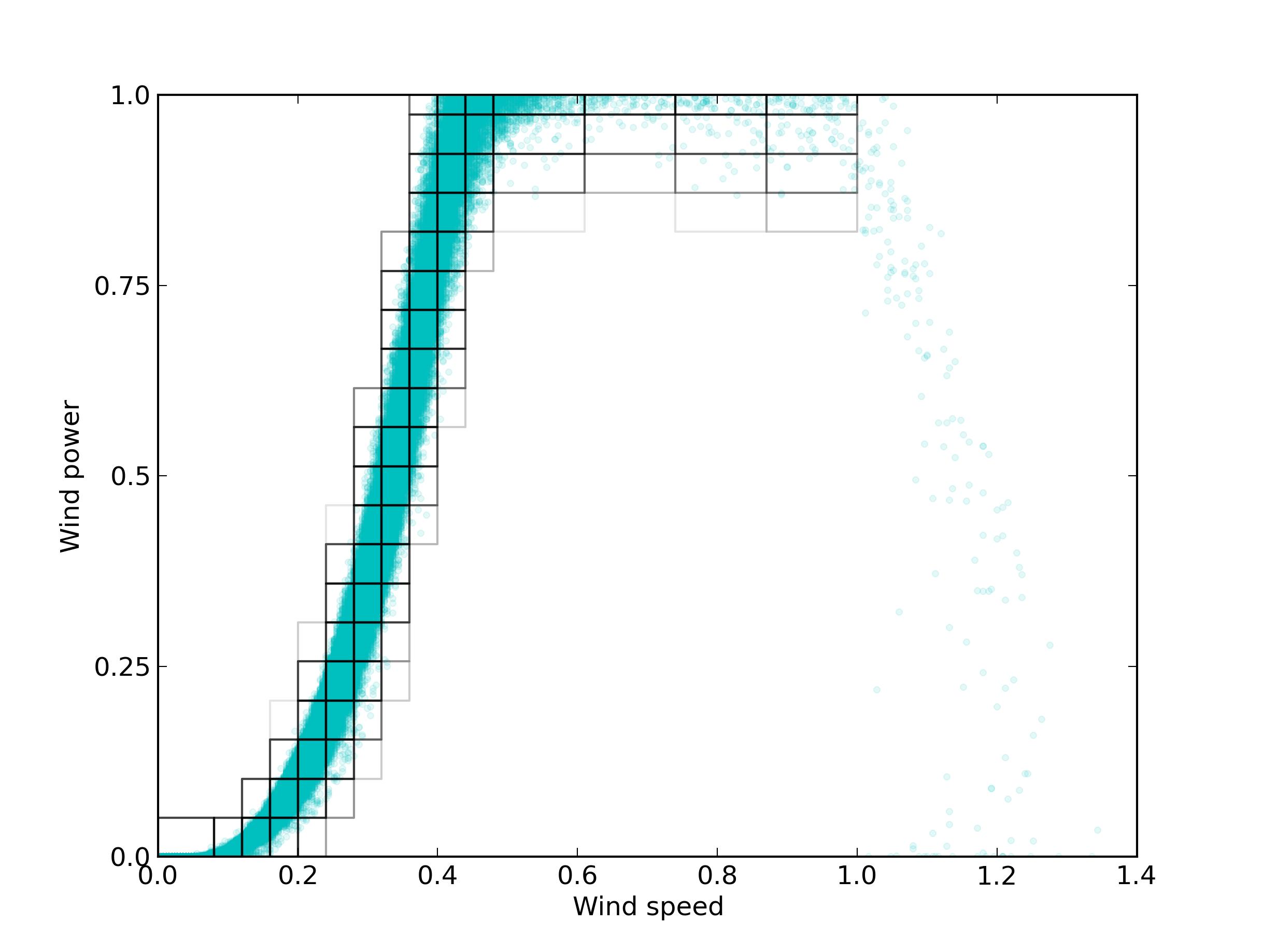}%
        \includegraphics[width = 0.5\textwidth]{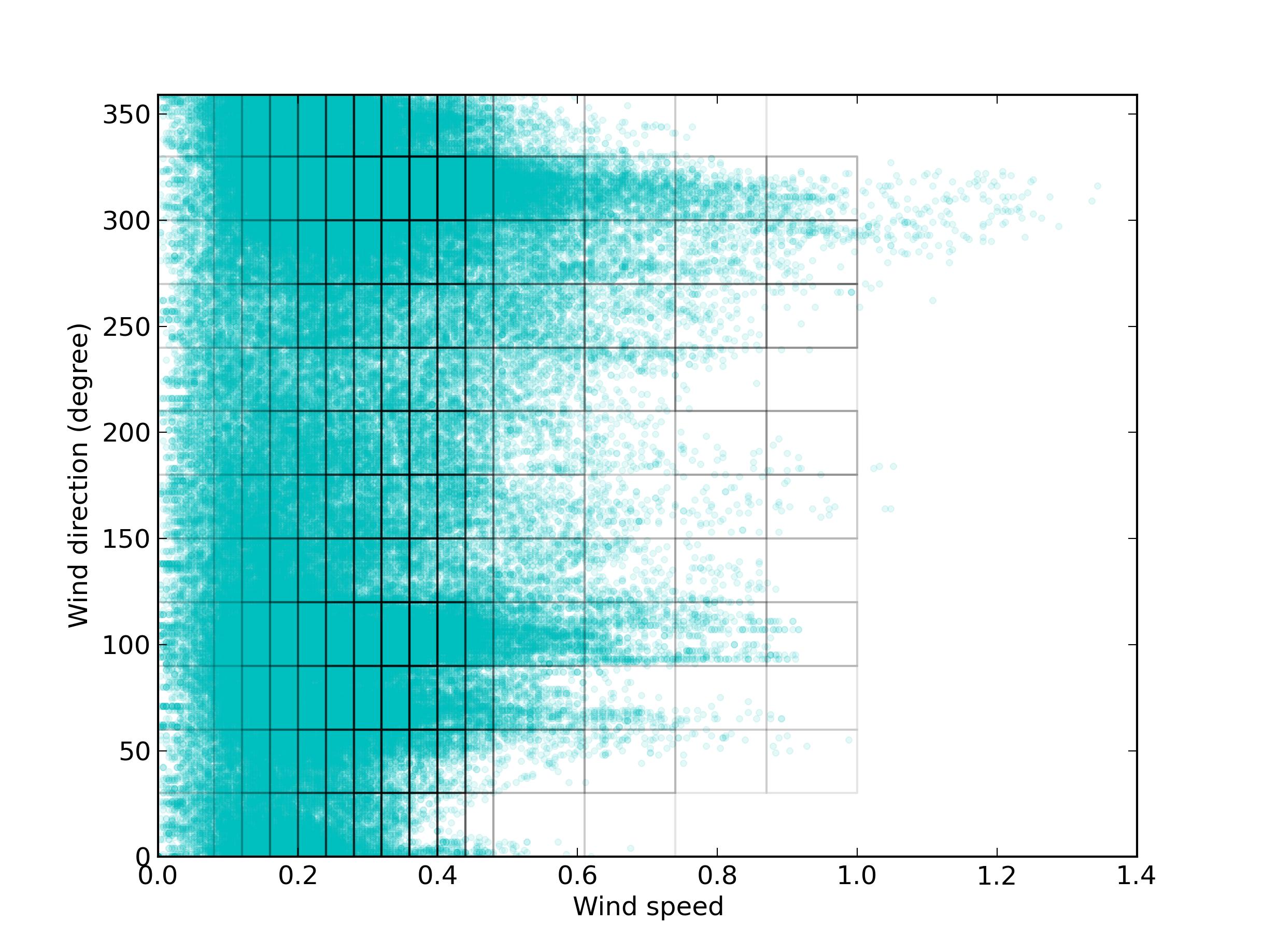}
        \caption{Representation of all data points projected into the: a) wind direction and speed plane (left); and, b) wind power and speed plane (right). Each rectangle is the projection of a state polyhedron into the two planes. Overall, they define the final state partition for the three-dimensional variable space.}
    \label{fig1}
    \end{figure}

    For this Markov model, each state is defined by the values of all three variables, namely the wind speed, wind direction and power output. Figure \ref{fig1} shows the data observations and the state partitions projected into the wind direction and speed plane (right) and the wind power and speed plane (left). 
    As expected, the observations projected into the wind power and speed plane define the characteristic power curve of the wind turbine.

    The data space is discretized unevenly to get a good resolution of the high-slope region of the power curve. In a previous work \cite{Lopes2012}, this partition was used in a time-homogeneous Markov chain and proved to lead to an accurate representation of the original data. The wind direction and power are divided by an equally spaced grid leading to 12 and 20 classes, respectively. The wind speed is divided as follows: values below the cut-in speed define one class; between the cut-in and rated wind speed the discretization is narrowed by selecting 10 classes ; and between the rated and cut-out wind speed discretization is widened and 4 classes are defined. Data points with wind speed above the cut-out wind speed are discarded. The complete state set is constructed by listing all possible combinations of the classes of each variable. Due to physical constraints between the variables, most of the states are empty and can are therefore discarded. This reduces the number of states from 3840 to 778, for this turbine.

    To compare the model with the original data, wind power, speed and direction time series were simulated adapting the method described by Sahin and Sen \cite{Sahin2001} to the cyclic time-inhomogeneous Markov model as follows.
    First, we compute the cumulative probability transition matrices $P^{\hbox{cum}}_r$ with entries $[P^{\hbox{cum}}_r]_{i,j} = \sum_{j^{\prime} = 0}^{j}p_{i,j^{\prime}}(r)$ . Then an initial state $s_{i}$, i.e. $X_0 = s_i$, is randomly selected. A new datapoint $X_{t+1}$ is generated by uniformly selecting a random number $\epsilon$ between zero and one and choosing for $X_{t+1}$ the corresponding state $s_{i^{\prime}}$ such that the probability of reaching it from the current state $s_i$ fullfils $[P_{r_t}^{\hbox{cum}}]_{i, i^{\prime}} \geq \epsilon$.
    Based on this discrete state sequence, a real value for the wind power/speed/direction variables is generated by sampling each state partition uniformly.
    
    \begin{figure}[htb]
        \centering
        \includegraphics[width = 0.32\textwidth]{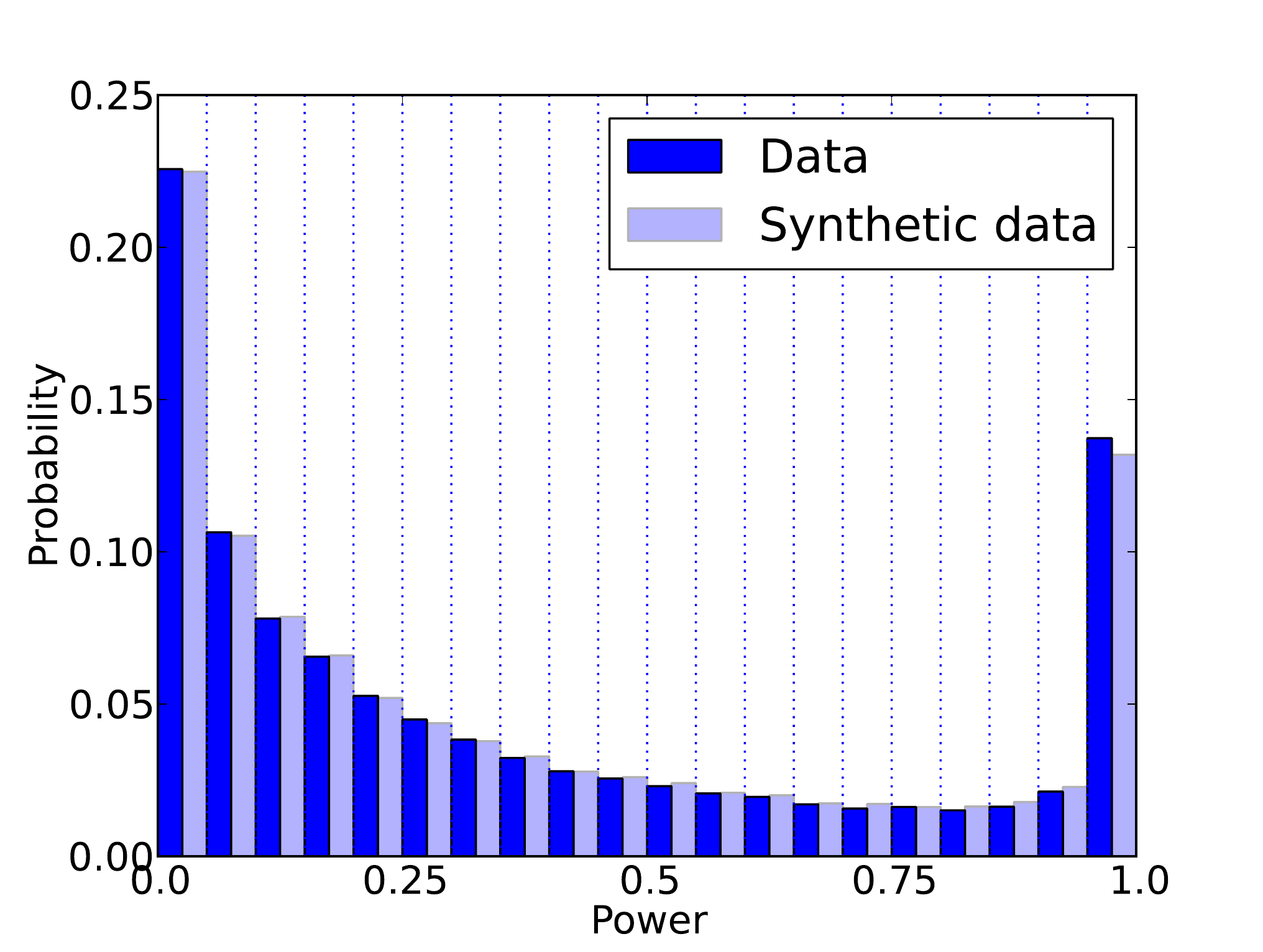}
        \includegraphics[width = 0.32\textwidth]{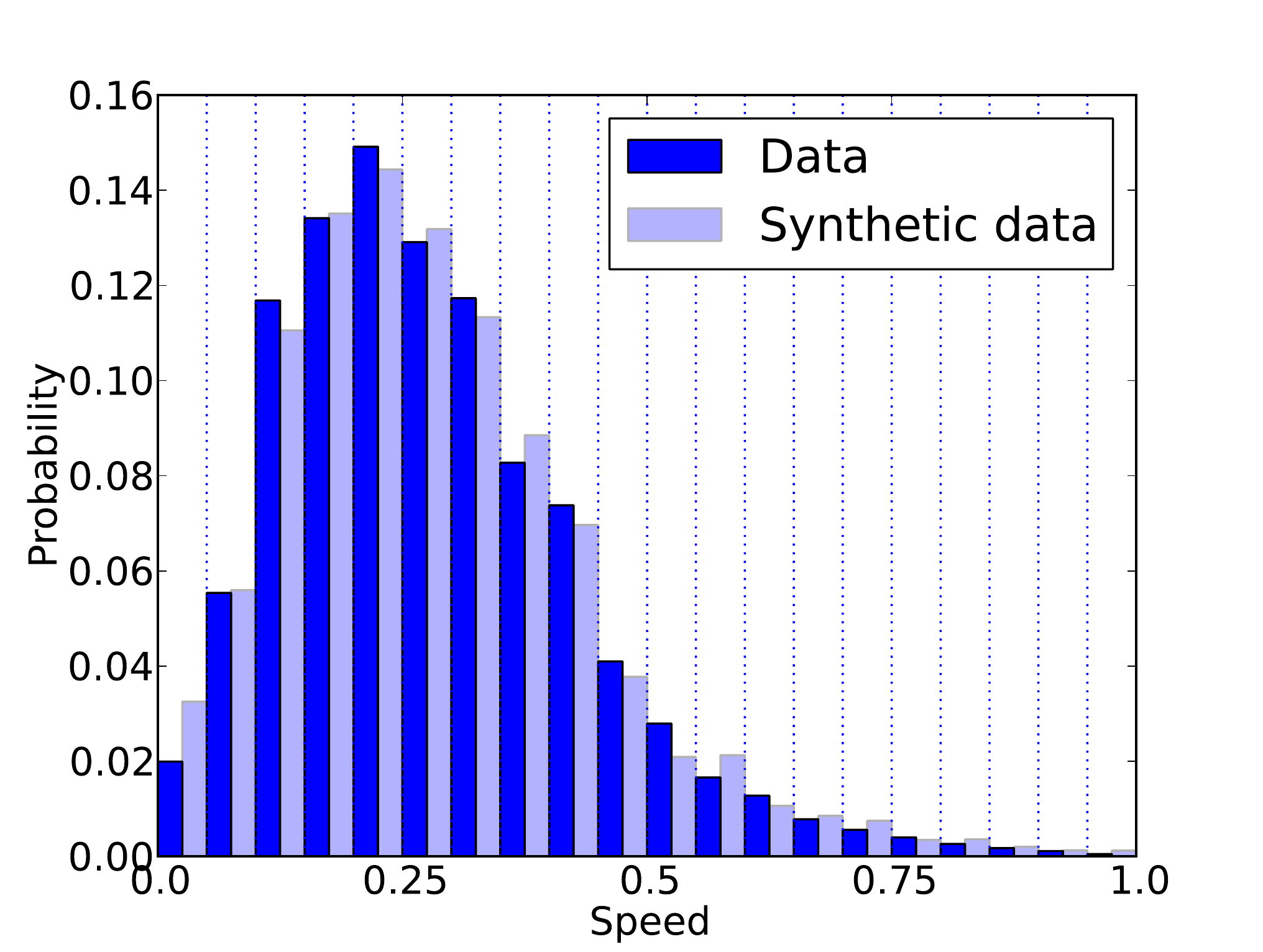}
        \includegraphics[width = 0.32\textwidth]{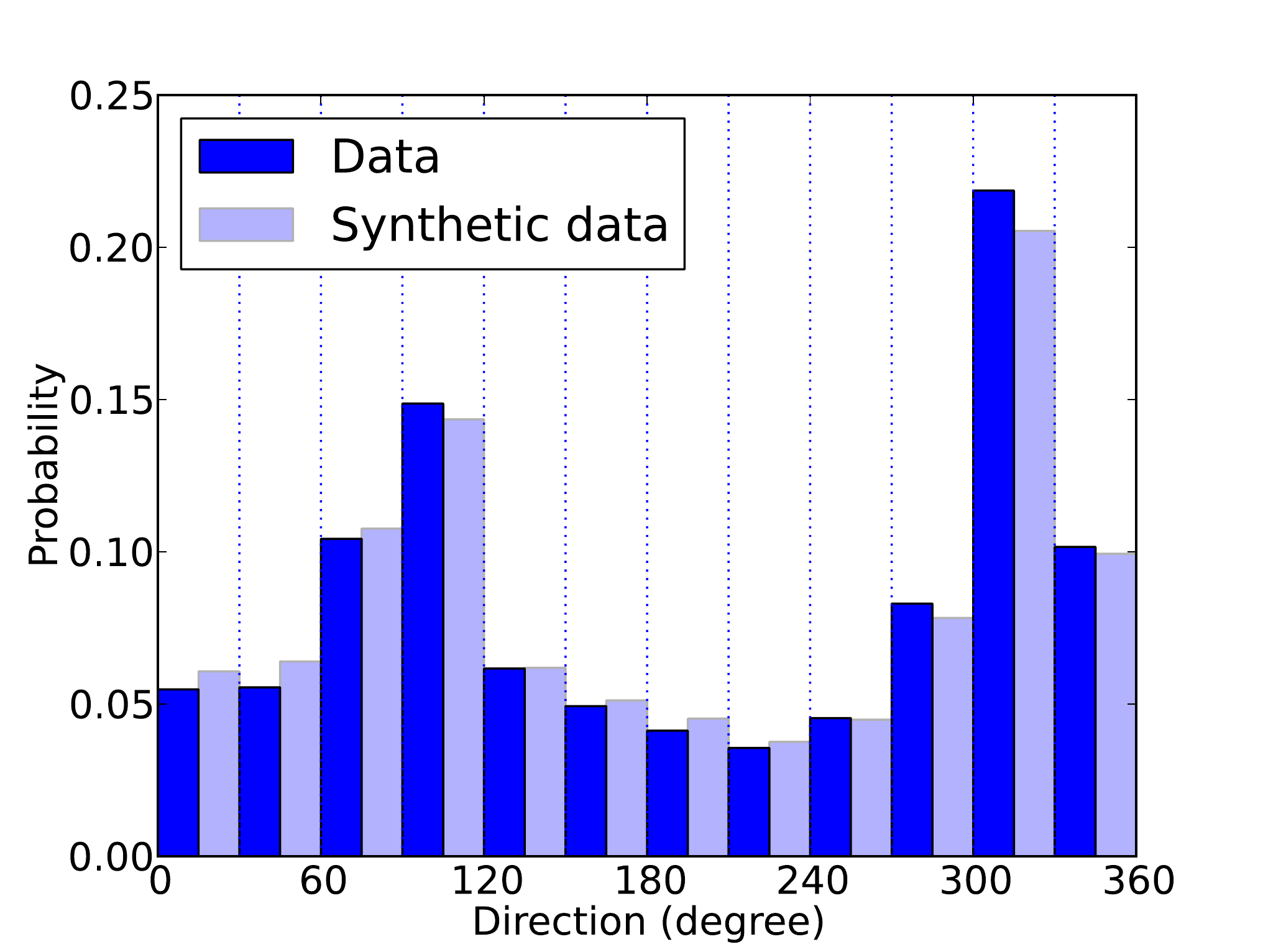}
        \caption{Comparison of the probability distribution of wind power (left), wind speed (middle) and wind direction (right) of the original with the synthesized data.}
    \label{fig2}
    \end{figure}

    \begin{figure}[htb]
        \centering
        \includegraphics[width = 0.45\textwidth]{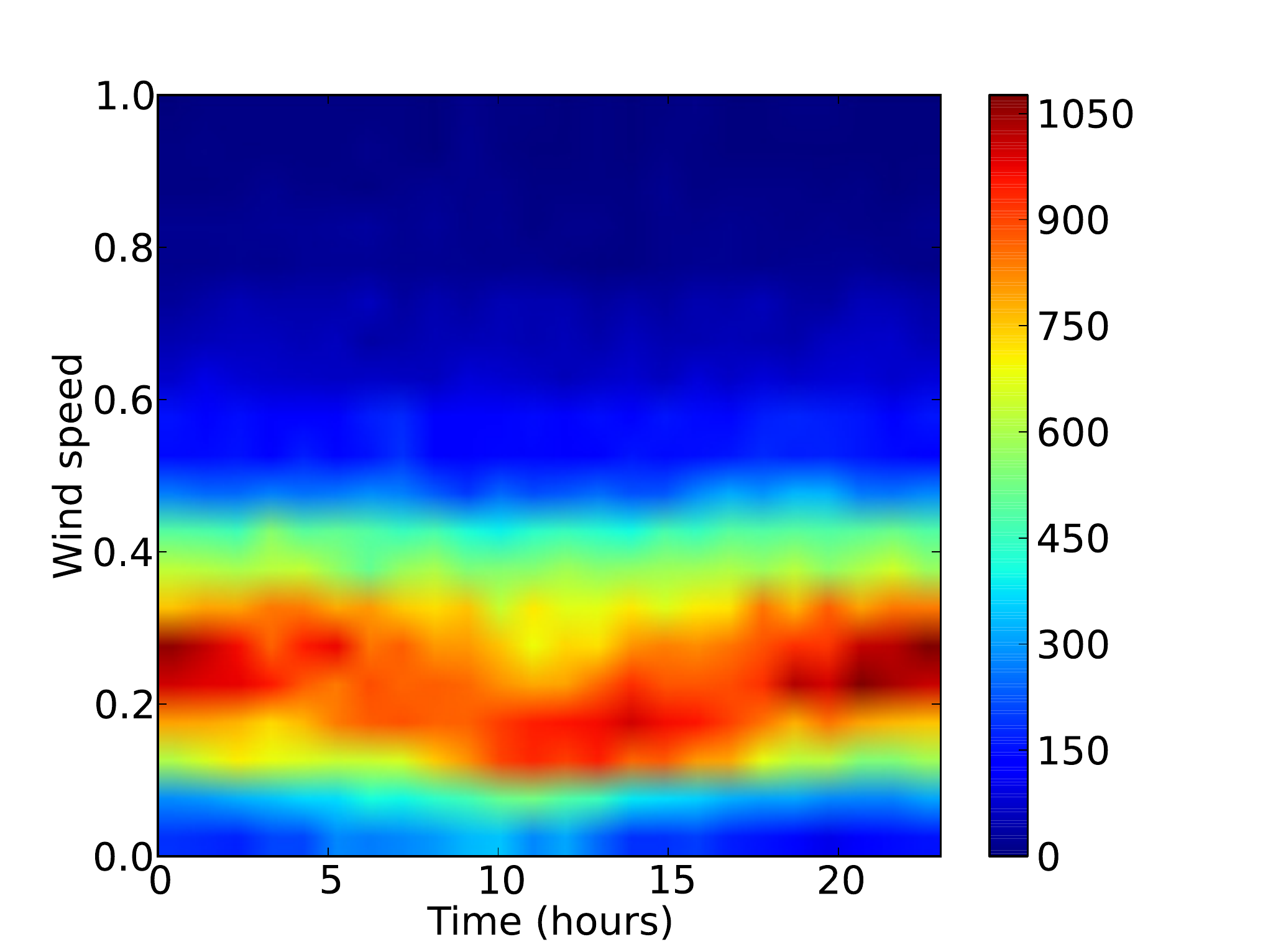}
        \includegraphics[width = 0.45\textwidth]{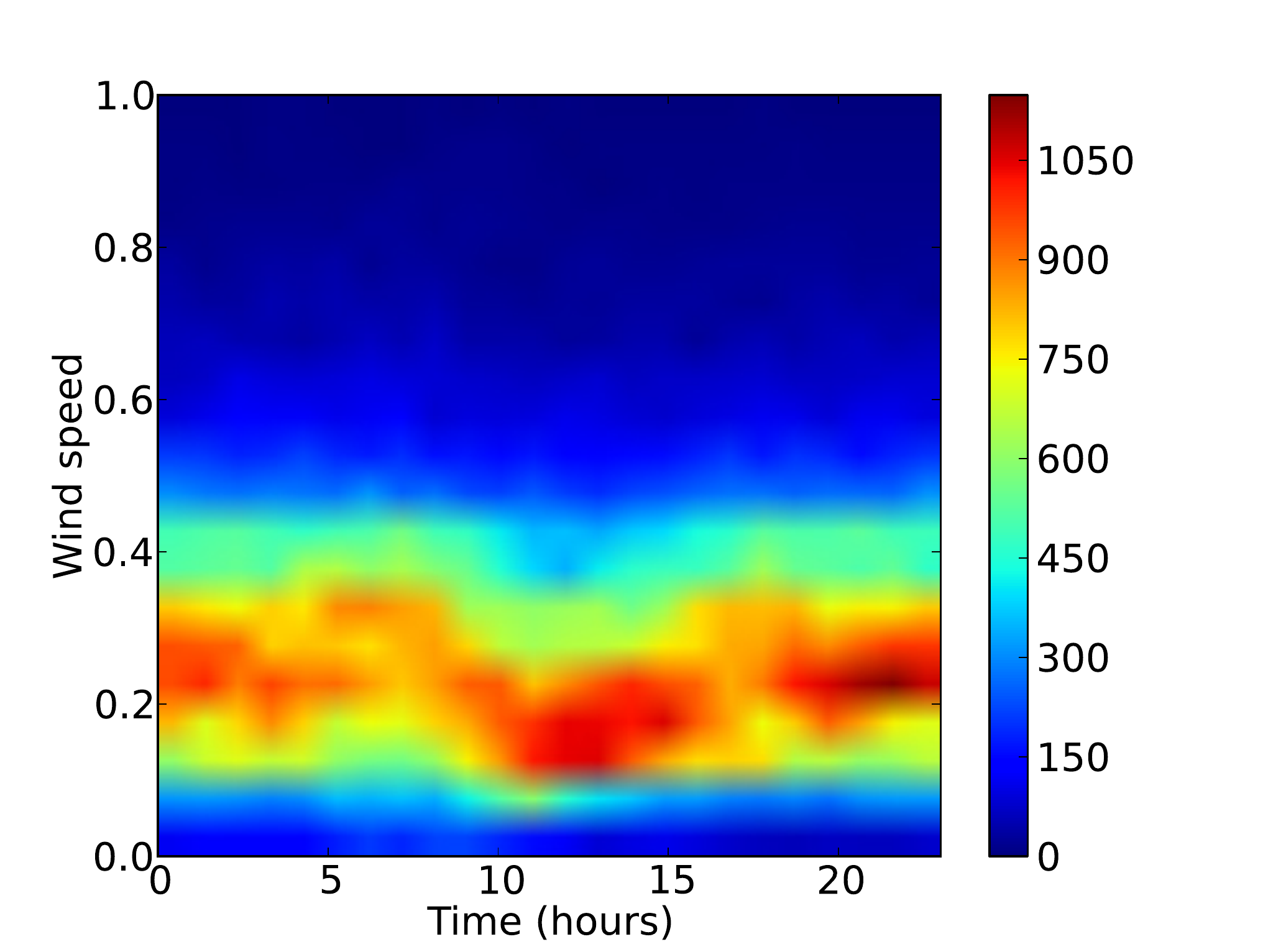}
        \includegraphics[width = 0.45\textwidth]{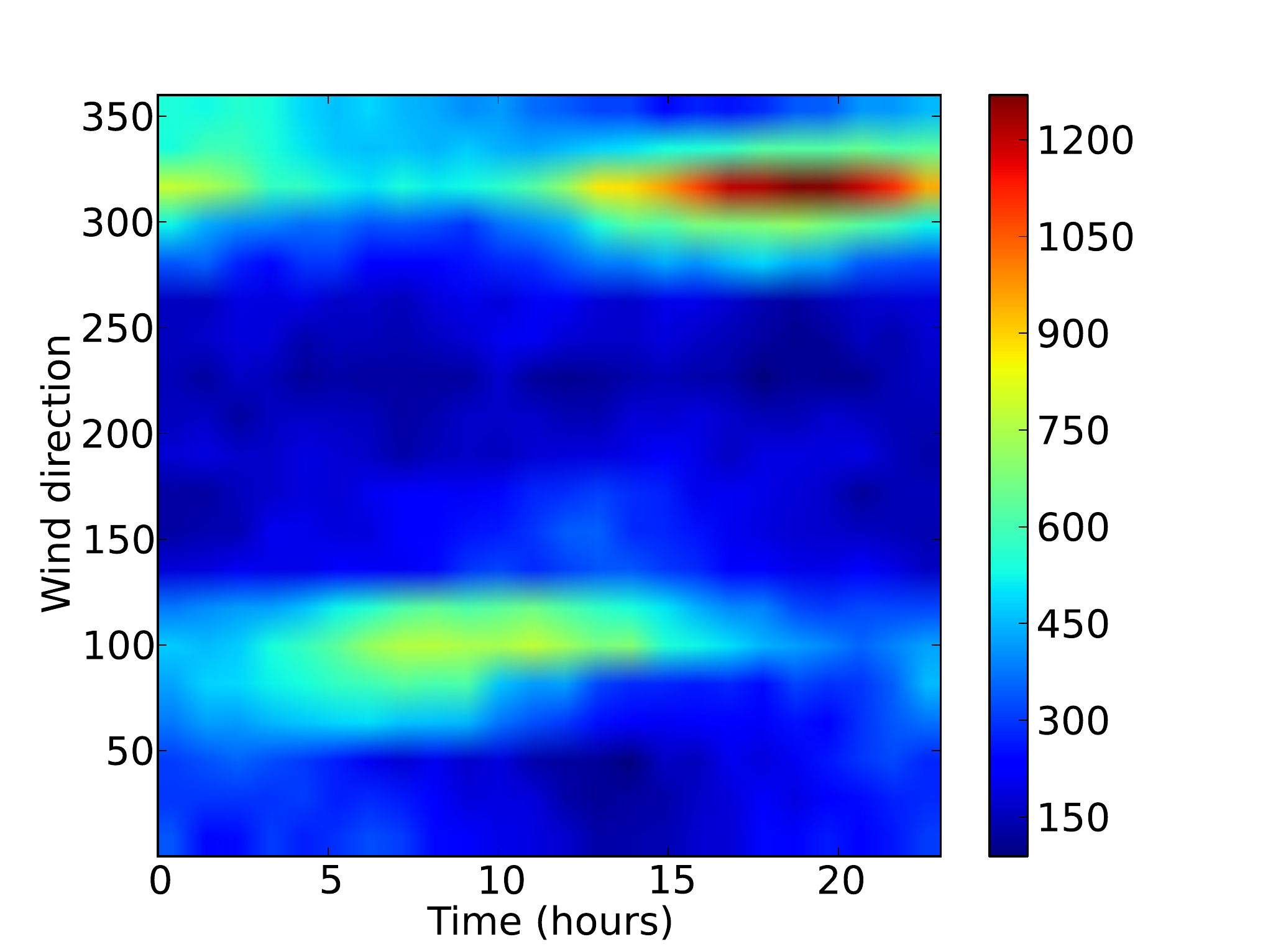}
        \includegraphics[width = 0.45\textwidth]{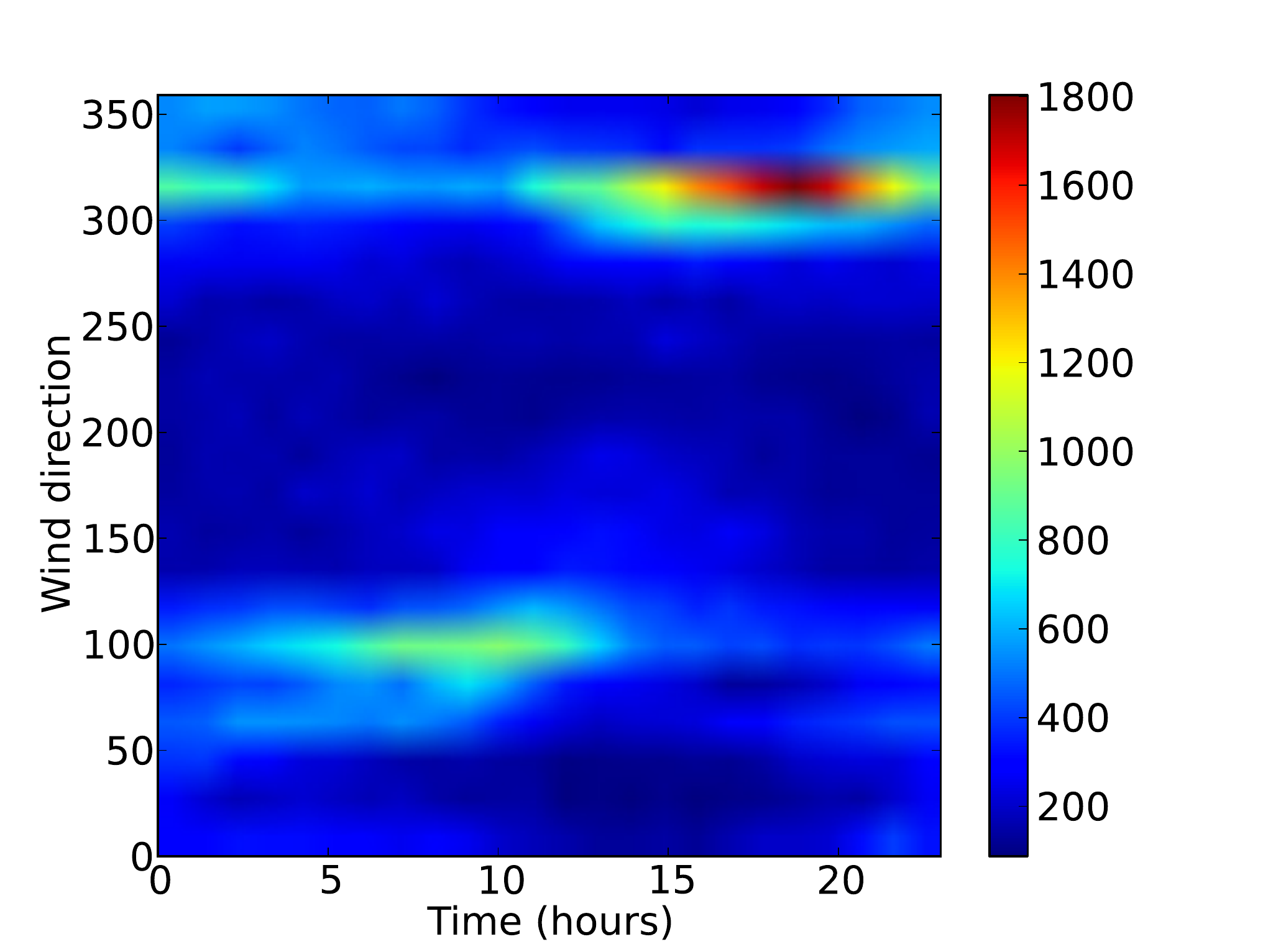}
        \caption{Two dimensional histograms of the synthetic time-series data, generated with the time-variant Markov model (left) and the original data (right): speed-time (top) and direction-time (bottom).}
    \label{fig3}
    \end{figure}
    
    Figures~\ref{fig2} and ~\ref{fig3} compare the original data with the synthesized data and demonstrate, that the model can capture the data's long-term statistics (fig.~\ref{fig2}) as well as the daily patterns (fig.~\ref{fig3}).
    
\section{The evolution of drift and diffusion in wind power output}
\label{sec:D1D2}

    With the procedure outlined in Sec.~\ref{sec:methods} and having the $144$ transition matrices generated as described in Sec.~\ref{sec:data} and \ref{section:MarkovProcess}, we can now reconstruct the time-dependent stochastic process by calculating the KM coefficients $\mathbf{D}^{(i)}(\mathbf{\hat{X}},t)$ at each time step $t=1,\dots,144$. Although we obtain the KM coefficients as a function of all three stochastic variables, $[P,v,\theta]$,
    we here consider only their dependency on the velocity and power production, $\bf{\hat{X}} = [P,v]$, 
    averaging over the contributions from $\theta$. 

\begin{figure}[H]
    \centering
    \includegraphics[width = 0.99\textwidth]{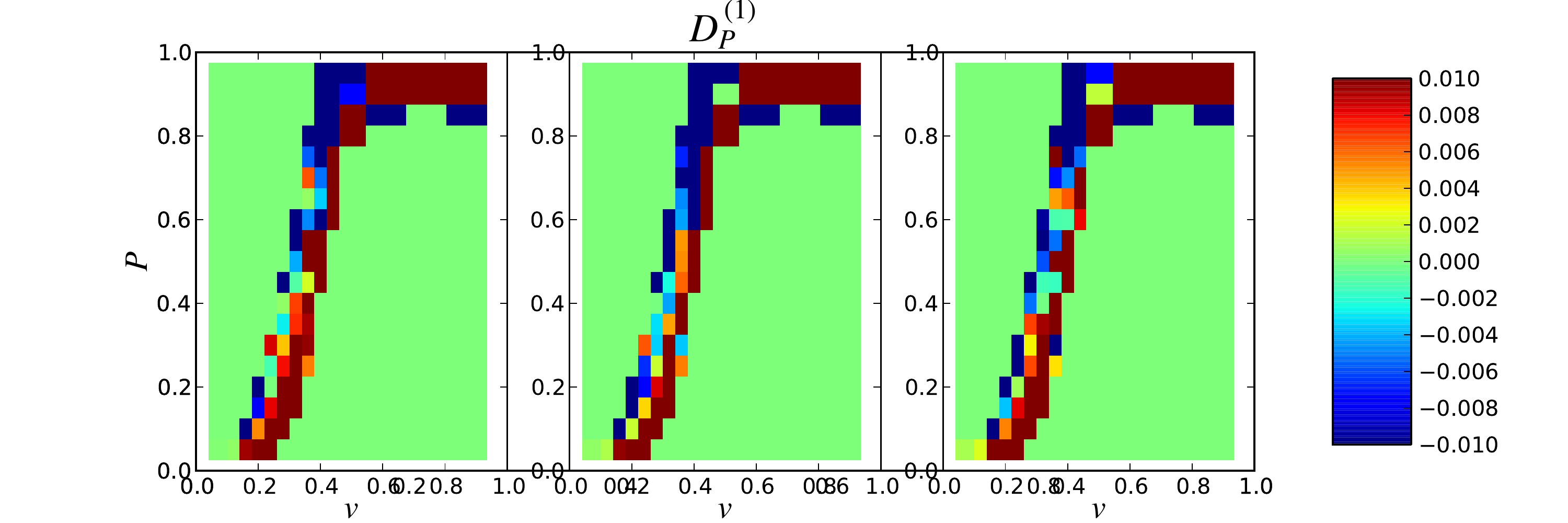}\\
    \includegraphics[width = 0.99\textwidth]{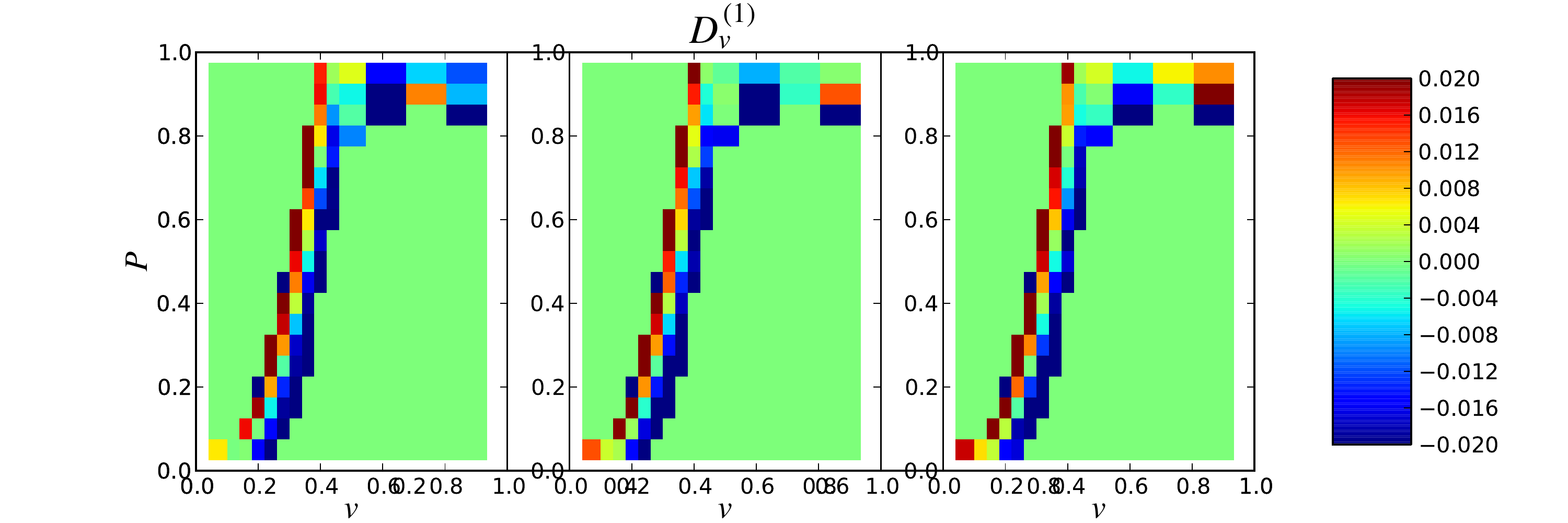}\\
    \includegraphics[width = 0.99\textwidth]{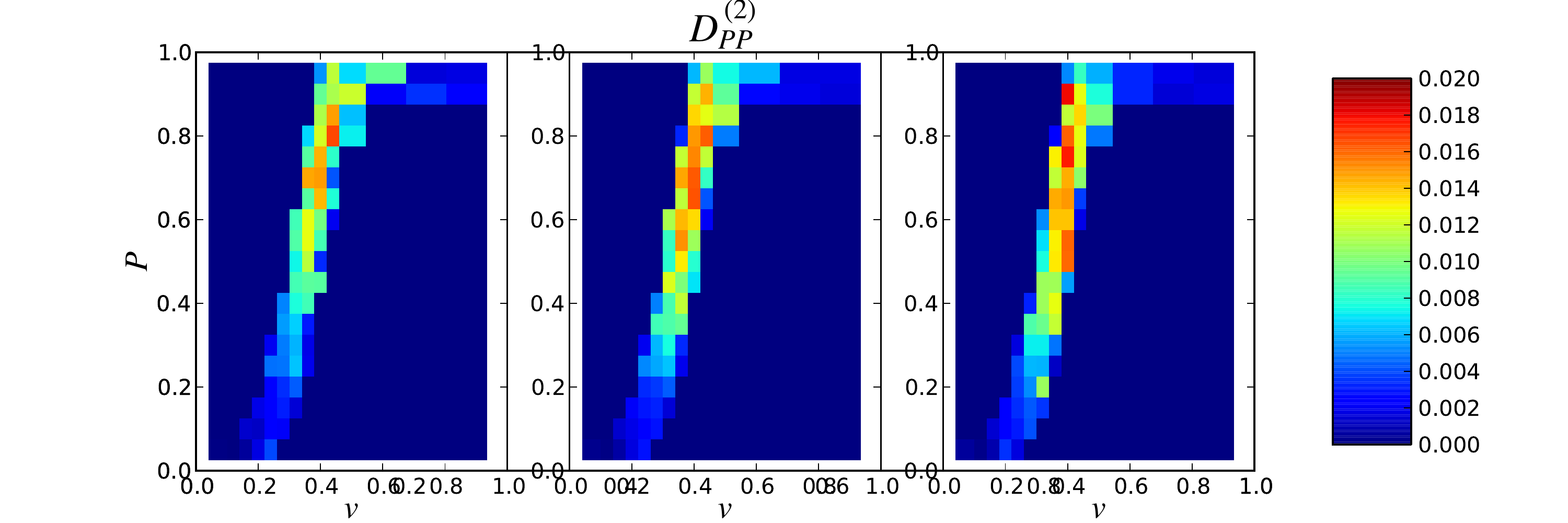}\\
    \includegraphics[width = 0.99\textwidth]{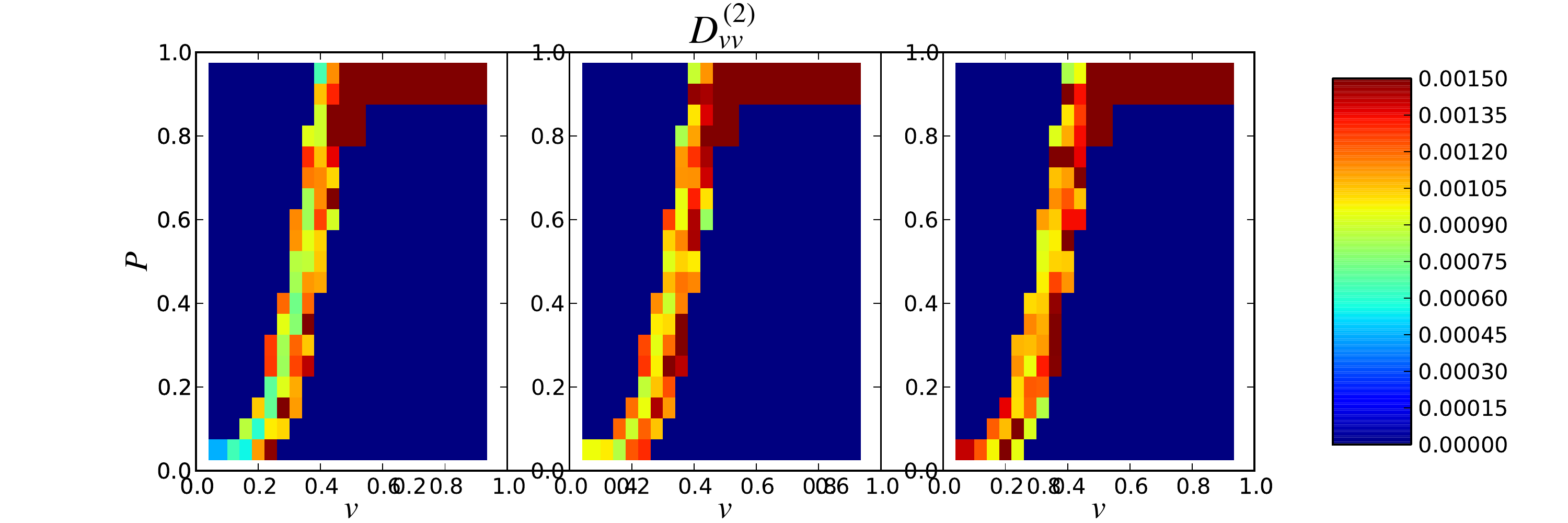}\\
    \includegraphics[width = 0.99\textwidth]{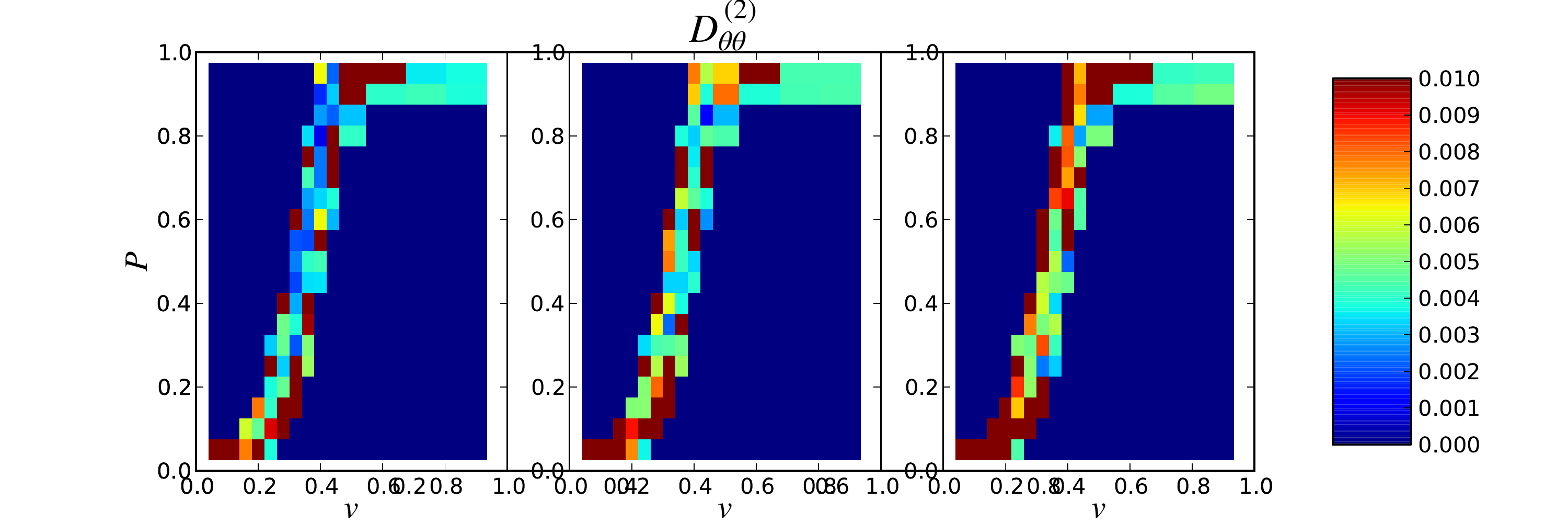}\\
    \caption{\protect
            The first (top two rows) and second (bottom three rows) Kramers-Moyal coefficients for various times (left: 6hours, middle: 12hours, right: 18hours).}
    \label{fig4}
    \end{figure}

\begin{figure}[H]
    \centering
    \includegraphics[width = 0.49\textwidth]{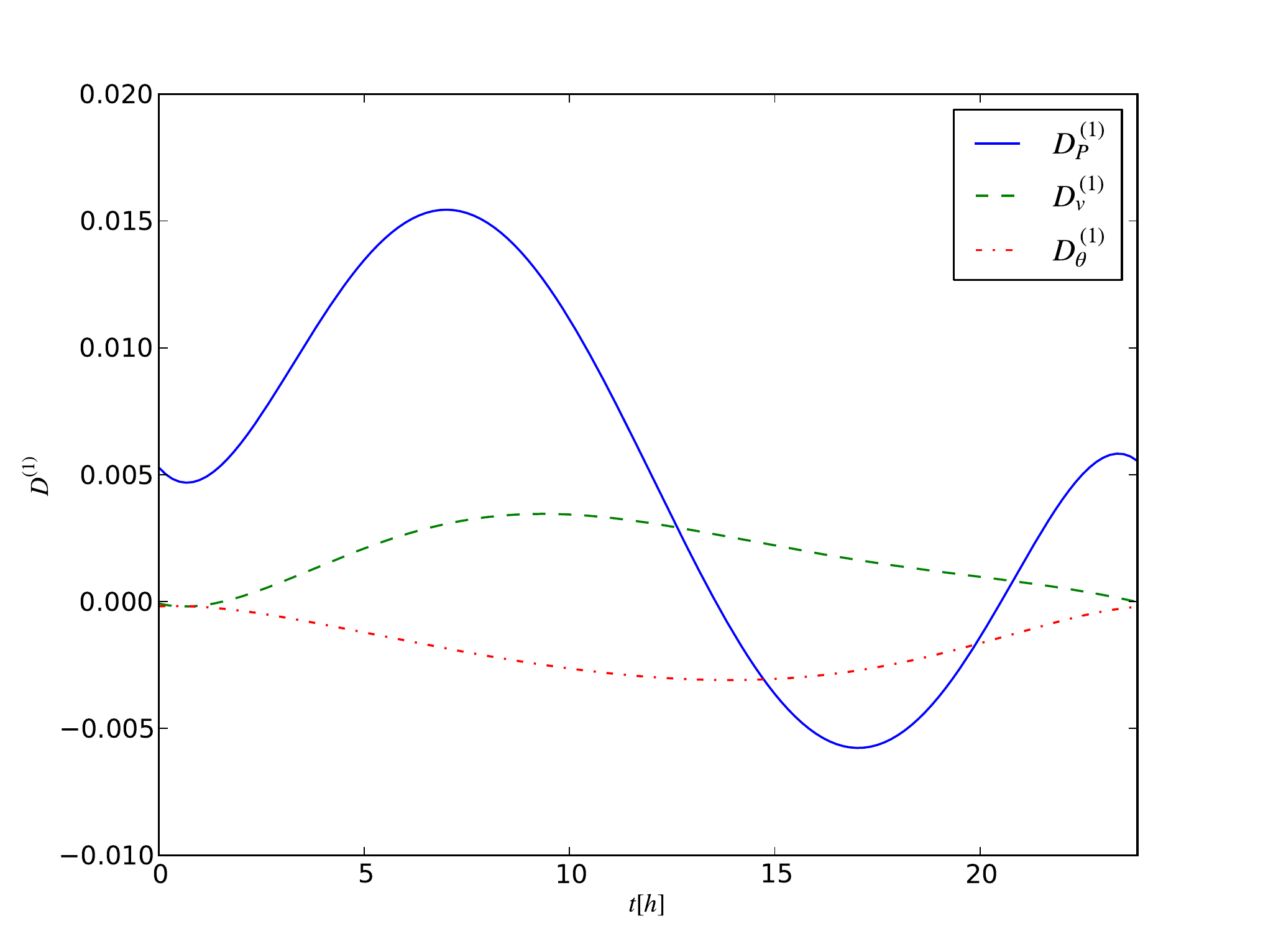}%
    \includegraphics[width = 0.49\textwidth]{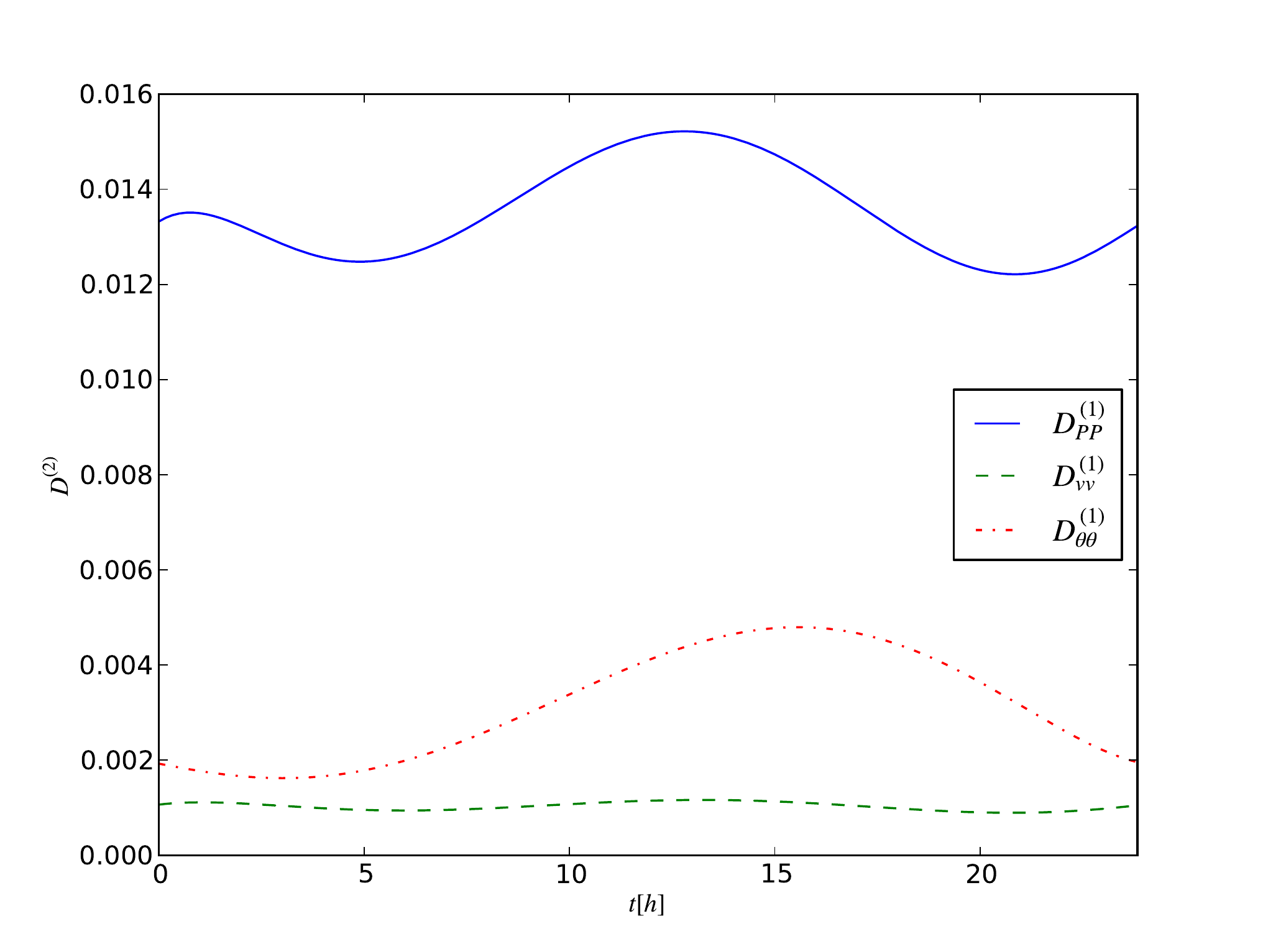}\\
    \caption{The first (left) and second (right) Kramers-Moyal coefficient, by components,  near the center of the power curve, $(v,P) = (0.34, 0.53)$,as a function of time.}
    \label{fig5}
    \end{figure}

    The results of this process are presented in Fig.\ref{fig4}, where the reconstructed KM coefficients are plotted for three time steps, namely at $6$, $12$ and $18$ hours. The support of the coefficients is limited to the available data which follows the power curves in the $v$-$P$ plane. From the inspection of Figs.~\ref{fig4}, changes in time seem not significative. This means, that even though both the Markov and the stochastic evolution model contain additional degrees of freedom due to their time-dependent formulation, they are capable of capturing the $v$-$P$-dependency, which is invariant. This is expected since the wind turbine operation characteristics should not change through the daily cycle. However, it can be seen in fig.~\ref{fig5} that the procedure is capable of detecting even subtle temporal changes in the transition matrix, which lead to strong daily changes in the KM coefficients.
    
    For all plotted times, the drift coefficients $D^{(1)}$ indicate a restoring force towards the power curve, in accordance with previous results\cite{Raischel2013}. The diffusion coefficients---only the diagonal components are shown here---show a order of magnitude weaker diffusion in the velocity than in the power, where the latter shows a strong component for diffusion in the $P-$ direction for high values of $v,P$. These results again are consistent with our previous analysis of a time-homogeneous model \cite{Raischel2013}. Remarkably, the out-of the $v,P-$ plane diffusion of the direction component $D^{(2)}_{\theta \theta}$ is strongest for both very high and very low velocities, and for intermediate velocities off the power curve.
    
    Next, we present a closer inspection of the time-dependence of both drift and diffusion, by considering their temporal evolution at a specific point, namely at $(v,P) = (0.34, 0.53)$, which is close to the center of the power curve. 
    Apparently, our method creates smooth curves for the temporal evolution. 
    This is expected since, as a consequence of the parametrization of the Markov model, it can be shown that the conditional moments used to derive the Drift and Diffusion coefficients also can be expressed by Bernstein polynomials in time. Most strikingly, it can be seen that the temporal evolution of both the drift and diffusion coefficient is decoupled between the components. Furthermore, for the same component the evolution of the diffusion coefficient seems to be delayed with respect to the drift. The dominant component is always the power production $P$, whose drift  changes from a positive maximum at $6$ h to a negative minimum at $17$ h, i.e.~the restoring force oscillates from a tendency to higher $P$ values in the morning to a tendency to lower $P$ values in the evening.     

    It should be noted that the chosen point $(v,P)$ is not necessarily characteristic of the wind field or of the turbine's power production. Other points along the power curve, specifically for low velocities and near the rated wind speed are either more frequent or more characteristic, and their analysis should give increased insight into the temporal evolution of wind speed and power production.


\section{Conclusions}
\label{sec:conclusions}

    We have shown in this paper how a time-dependent multi-dimensional stochastic process can be reconstructed from experimental data. Our method provides results in terms of the time-dependent transition matrix of a Markov model, from which the time-dependent Kramers-Moyal coefficients for a corresponding continuous process can be calculated. Application of this method to data from a turbine in a wind park gives results consistent with a previous time-independent method, and adds surprising new insight into the temporal dynamics of the wind field and the machine power production. Preliminary results have shown that the dependence with time observed in Fig. 5 changes depending which region of the power curve we choose. A more systematic study for the full power-velocity range will be carried out in an extended study. Future research will also address the question of applicability of our method to more general cases, dealing also with the reliability and relative errors of this approach.  
    
    The aforementioned equivalence of the transition Matrix and the KM coefficients is valid if two requirements are fulfilled. First, the transition amplitudes need to have Gaussian shape, which corresponds to the existence of Gaussian noise in the stochastic process. The validity of this assumption has been checked previously for a similar system and can be reasonably assumed in this case. Secondly, the binning of the stochastic variables for determining the Markov process transition matrix must be small enough\cite{vankampen}. We will investigate the validity of this assumption and the corresponding errors in a forthcoming publication.

\section*{Acknowledgements}
    The authors acknowledge helpful discussions with David Kleinhans, who provided the basic idea to calculate the Kramers-Moyal coefficients directly.
    The authors thank Funda\c{c}\~ao para a Ci\^encia e a Tecnologia for financial support under PEst-OE/FIS/UI0618/2011, PEst-OE/MAT/UI0152/2011, FCOMP-01-0124-FEDER-016080 and SFRH/BD/86934/2012 (TS). VVL thanks the Prometeo Project of SENESCYT (Ecuador) for financial support. PGL thanks the German Environment Ministry for financial support (0325577B). This work is part of a bilateral cooperation DRI/DAAD/1208/2013
    supported by FCT and Deutscher Akademischer Auslandsdienst (DAAD). 
    FR assisted in fundamental research in the frame of T\'{A}MOP 4.2.4. A/2-11-1-2012-0001 National Excellence Program – Elaborating and operating an inland student and researcher personal support system, was realised with personal support. The project was subsidized by the European Union and co-financed by the European Social Fund. FR would like to thank F.Kun, Univ. Debrecen, for his hospitality.

\end{document}